\newcommand{\be}{\begin{equation}}
\newcommand{\ee}{\end{equation}}
\def\bea{\begin{eqnarray}}
\def\eea{\end{eqnarray}}
\documentstyle[twoside,fleqn,espcrc2]{article}
\title{ From Neutrino Masses to Proton Decay}

\author{P. Ramond
\address{Institute for Fundamental Theory\\
Department of Physics,\\ University of Florida\\ 
Gainesville, Fl 32611}}

\begin{document}
\begin{abstract}
\noindent Current theoretical and experimental issues are reviewed in
the light of the recent SuperKamiokande discovery.  By using
quark-lepton symmetries, derived from  Grand Unification and/or string
theories, we show how to determine the  necessary neutrino 
parameters. In addition,  the seesaw neutrino masses   set the scale
for  the proton decay operators by ``measuring'' the standard model
cut-off. The SuperKamiokande values suggest  that proton decay is
likely to be observed early in the XXIst Century.

\end{abstract}
\maketitle

\section{Neutrino Story}
Once it became apparent that the spectrum of $\beta$ electrons was 
continuous~\cite{CHAD,ELWO}, something drastic had to be done! 
 In December 1930,  in a  letter that starts with 
typical panache, $``${\it  Dear Radioactive Ladies and Gentlemen...}", 
W. Pauli puts forward a ``{\it desperate}" way out: there is a
companion neutral particle to the $\beta$ electron. Thus earthlings became aware of
the {\it neutrino}, so named in 
1933 by Fermi (Pauli's original name, {\it neutron},
superseded by Chadwick's discovery of a heavy neutral particle), 
implying that there is something small about it, specifically 
its  mass, although nobody at that time thought it was {\it that} small. 

Fifteen years later, B. Pontecorvo~\cite{PONTA}
proposes the unthinkable, that neutrinos can be detected:  an electron 
neutrino that hits a ${^{37}Cl}$ atom will transform it into the inert 
radioactive gas ${^{37}Ar}$, which  can be stored and then
detected through  radioactive decay.  Pontecorvo did not publish
the report, perhaps because of the times, or because 
Fermi  thought the idea ingenious but not immediately  relevant.

In 1956, using a scintillation counter experiment they had proposed 
three years earlier~\cite{COREA},  Cowan and Reines~\cite{COREB} 
discover electron antineutrinos through the reaction
$\overline \nu_e+p\rightarrow e^+ +n$. Cowan passed away before 1995,
the year Fred Reines was awarded the Nobel Prize for their discovery. 
There emerge two  lessons in neutrino physics: not only is patience
required but also longevity: it took $26$ years from birth to
detection and then another $39$ for the Nobel Committee to recognize 
the achievement! This should encourage   physicists to 
train their children at the earliest age to follow their footsteps at 
the earliest possible age, in order to establish dynasties of neutrino 
physicists.  Perhaps then Nobel prizes will be awarded to scientific families? 

In 1956, it was rumored that  Davis~\cite{DAVISA},  following  Pontecorvo's 
proposal,  had found evidence for neutrinos coming from a pile, and 
Pontecorvo~\cite{PONTB}, influenced by the recent work of   Gell-Mann 
and Pais, theorized that  an antineutrino produced in the Savannah reactor 
could oscillate into a neutrino and be detected. The rumor went
away, but the idea of neutrino oscillations was born; it has remained
with us ever since. 

Neutrinos  give up their secrets very grudgingly: its
helicity was measured in 1958 by M. Goldhaber~\cite{MGOLD},
but it took 40 more years for experimentalists to produce convincing 
evidence  for its mass. The second neutrino, the muon neutrino  is 
detected~\cite{2NEUT} in 1962, (long anticipated by theorists Inou\"e 
and Sakata in 1943~\cite{INSA}). This time things went a bit faster as 
it took only 19 years from theory (1943) to discovery (1962) and 26 years 
to Nobel recognition (1988). 

That same year, Maki, Nakagawa and Sakata~\cite{MANASA} introduce 
two crucial ideas:   neutrino flavors can mix, and their  
 mixing can cause one type of neutrino to oscillate 
into the other (called today flavor oscillation). This is possible 
only if the two neutrino flavors have different masses. 

In 1964, using Bahcall's result~\cite{BAH} of an enhanced capture 
rate of ${^8B}$ neutrinos through an excited state of ${^{37}Ar}$, Davis~\cite{DAVISB}
proposes to search for ${^8B}$ solar neutrinos using a $100,000$ gallon
tank of cleaning fluid deep underground.  Soon after, R. Davis starts 
his epochal experiment at the Homestake mine, marking the
beginning of the solar neutrino watch which continues to this day. In
1968, Davis et al reported~\cite{DAVISC} a deficit in the solar
neutrino flux, 
a result that stands to this day as a truly
remarkable experimental {\it tour de force}. Shortly after, Gribov and
Pontecorvo~\cite{GRIPO} interpreted the deficit as evidence for neutrino oscillations.

In the early 1970's, with the idea of quark-lepton
symmetries~\cite{PASA,GG} suggests that the proton could
be unstable. This brings about the construction of  underground 
 detectors, large enough to monitor many protons, and instrumentalized to 
detect the \v Cerenkov light emitted by its decay products. By the middle 1980's,
several such detectors are in place. They fail to detect proton decay, 
but in a remarkable serendipitous turn of events, 150,000 years earlier, a 
supernova erupted in the large Magellanic Cloud, and in 1987, 
its burst of neutrinos was detected in these detectors! All of 
a sudden, proton decay detectors turn their attention to neutrinos, 
while to this day still waiting for its protons to decay! Today, 
these detectors have shown great success in measuring the effects of 
solar and atmospheric neutrinos. They continue their unheralded
watch for signs of proton decay, reassured in the knowledge 
that lepton number and baryon number violations are connected 
in most theories, leading to correlations between neutrino 
masses and  proton decay rates.

\section{Standard Model Neutrinos}
The standard model of electro-weak and strong interactions contains 
three left-handed neutrinos.  The three neutrinos are represented by 
two-components Weyl spinors, $\nu^{}_{i}$, $i=e,\mu,\tau$, each 
describing a left-handed fermion (right-handed antifermion). As 
the upper components of weak isodoublets $L^{}_i$, they have 
$I^{}_{3W}=1/2$, and a unit of the global $i$th lepton number. 

These standard model neutrinos are strictly massless. The only 
Lorentz scalar made out of these neutrinos is the Majorana mass, 
of the form
$\nu^{t}_{i}\nu^{}_{j}$; it has the quantum numbers of a weak 
isotriplet, with third component  $I^{}_{3W}=1$, as well as two 
units of total lepton number.  Higgs isotriplet with two units 
 of lepton number could generate neutrino Majorana masses, but 
there is no such higgs in the Standard Model:  there are no 
tree-level neutrino masses in the standard model.

Quantum corrections, however,  are  not limited to
renormalizable couplings, and it is easy to make a weak isotriplet out
of two isodoublets, yielding the $SU(2)\times U(1)$ invariant
$L^t_i\vec\tau L^{}_j\cdot H^t_{}\vec\tau H$, where $H$ is the Higgs
doublet. As  this term is not invariant under lepton number, it is 
not be generated in perturbation theory. Thus the
important conclusion: {\it The standard model neutrinos are kept
massless by global chiral lepton number symmetry}. The detection 
of  neutrino masses is therefore {\it a tangible indication 
of physics beyond the standard model}.

\section{Experimental Issues}
From the solar neutrino deficit to the spectacular result from
SuperKamiokande, experiments suggest that neutrinos have masses,
providing the first credible evidence for physics beyond the standard
model. As we stand at the end of this Century, there remains several  
burning issues in neutrino physics that can be settled by future experiments:

\begin{itemize}
\item The origin of the Solar Neutrino Deficit

This is currently being addressed by SuperK, in their measurement of the shape 
of the ${^8}B$ spectrum, of day-night asymmetry and of the seasonal 
variation of the neutrino flux. Their reach will soon be improved 
by lowering their threshold energy. 

SNO is  joining the hunt, and is expected to provide a more accurate 
measurement of the Boron flux. Its {\it raison d'\^etre}, however, 
is the ability to measure neutral current interactions. If there 
are no sterile neutrinos, we might have a flavor independent 
measurement of the solar neutrino flux, while measuring at the 
same time the electron neutrino flux!

This experiment will be joined by BOREXINO, designed to measure 
neutrinos from the $^7Be$ capture. These neutrinos are suppressed 
in the small angle MSW solution, which could explain the results 
from the $p-p$ solar neutrino experiments and those that measure the Boron neutrinos. 
\item Atmospheric Neutrinos

Here, there are several long baseline experiments to monitor muon 
neutrino beams and corroborate the SuperK results. The first, 
called K2K, already in progress, sends a beam from KEK to SuperK. 
Another, called MINOS, will monitor a FermiLab neutrino beam at 
the Soudan mine, 730 km away. A third experiment under consideration 
would send a CERN beam towards the Gran Sasso laboratory (also about 
730 km away!). Eventually, these experiments hope to detect the 
appearance of a tau neutrino.
\end{itemize}

This brief survey of upcoming experiments in neutrino physics is 
intended to give a flavor of things to come. These experiments will 
not only measure neutrino parameters (masses and mixing angles), 
but will help us answer fundamental questions about the nature of 
neutrinos. But the future of neutrino detectors may be even brighter. 
Many of us expect them to detect proton decay, thus realizing the   
kinship between leptons and quarks. There is even increasing talk 
of producing intense neutrino beams in muon storage rings, and 
at this workshop of building a mammoth proton decay/neutrino detector! 

\section{Neutrino Masses}
Neutrinos must be extraordinarily light: experiments indicate  $m_{\nu_e}< 
10~ {\rm eV}$, $m_{\nu_\mu}< 170~ {\rm keV}$, $m_{\nu_\tau}< 18~ {\rm
MeV}$~\cite{PDG}, and any model of neutrino masses must explain this
suppression. 

The natural way to generate neutrinos masses is to introduce for
each one its electroweak singlet Dirac partner, $\overline
N^{}_i$. These appear naturally in the Grand Unified group
$SO(10)$ where they complete each family into its spinor
representation. Neutrino Dirac masses will then be generated by the 
couplings $L^{}_i\overline N^{}_j H$ after electroweak breaking. 
However, unless there are extraordinary suppressions,   these
couplings generate masses that 
are way too big, of the same order of magnitude as the masses of the charged
elementary particles $m\sim\Delta I_w=1/2$.

Based on recent ideas from string theory, it has been
proposed~\cite{HA} that the world of four dimensions is in fact 
a ``brane" immersed in a higher dimensional space. In this view, 
all fields with electroweak quantum numbers live on the brane, 
while standard model singlet fields can live on the ``bulk" as 
well. One such field is the graviton, others could be the right-handed 
neutrinos. Their couplings to the brane are reduced by geometrical 
factors, and the smallness of neutrino masses is due to the 
naturally small coupling between brane and bulk fields.  

In the absence of any credible dynamics for the physics of the bulk, 
we think  that {\it ``one neutrino on the brane is worth two 
in the bulk"}. We take  the more conservative approach where  the 
bulk does opens up, but  at much shorter scales. One indication of such 
a scale is that at which the gauge couplings unify, the other is 
given by the value of neutrino masses.   This is achieved by  introducing
Majorana mass terms $\overline N^{}_i\overline N^{}_j$ for the
right-handed neutrinos. The masses of these new degrees of freedom are 
arbitrary, as they have no electroweak quantum numbers, $M\sim\Delta
I_w=0$. If they are much larger than the electroweak scale, the
neutrino masses are suppressed relative to that of their charged
counterparts by the ratio of the electroweak scale to that new scale: 
the mass matrix  (in $3\times 3$ block form) is
\be
\hskip 1in \pmatrix{0& m\cr m&M}\ ,
\ee
leading, for each family,  to one small and one large eigenvalue 
\be
m_\nu~\sim~ m\cdot {m\over M}~\sim~ \left(\Delta I_w={1\over 2}\right)\cdot 
\left({ \Delta I_w={1\over 2}
\over \Delta I_w=0 }\right)\ .\ee 
This seesaw mechanism~\cite{SEESAW} provides a natural
explanation for  small neutrino masses as long as lepton
number is broken at a large scale $M$. With $M$ around the energy at
which the gauge couplings unify, this yields neutrino masses at or
below tenths of eVs, consistent with the SuperK results. 

The lepton flavor mixing comes from  the diagonalization
of the charged lepton Yukawa couplings, and  of the neutrino
mass matrix. From the charged lepton Yukawas, we obtain ${\cal U}_e^{}$, 
the unitary matrix that rotates the
lepton doublets $L^{}_i$. From the neutrino Majorana matrix, we obtain
$\cal U_\nu$, the matrix that diagonalizes the Majorana mass matrix. 
The $6\times 6$ seesaw Majorana matrix can be written in $3\times 3$
block form
\be
{\cal M}={\cal V}_\nu^t ~{\cal D} {\cal V}^{}_\nu\sim\pmatrix {{\cal
U}_{\nu\nu}&\epsilon {\cal U}^{}_{\nu N}\cr
\epsilon{\cal U}^{t}_{N \nu}&{\cal U}^{}_{NN}\cr}\ ,\ee
where $\epsilon$ is the tiny ratio of the electroweak to lepton
number violating scales, and ${\cal D}={\rm diag}(\epsilon^2{\cal D}_\nu, {\cal D}_N)$,
 is a diagonal matrix. ${\cal D}_\nu$ contains the
three neutrino masses, and $\epsilon^2$ is the seesaw suppression. The
weak charged current is then given by
\be
j^+_\mu=e^\dagger_i\sigma_\mu {\cal U}^{ij}_{MNS}\nu_j\ ,\ee
where
\be
{\cal U}^{}_{MNS}={\cal U}^{}_e{\cal U}^\dagger_\nu\ ,\ee
is the Maki-Nakagawa-Sakata~\cite{MANASA} (MNS) flavor mixing matrix, 
the analog of the CKM matrix in the quark sector. 

In the seesaw-augmented standard model, this mixing matrix is totally 
arbitrary. It contains, as does the CKM matrix, three rotation angles,
and one CP-violating phase. In the seesaw scenario, it also contains
 two additional CP-violating phases
which cannot be absorbed in a redefinition of the neutrino
fields, because of their Majorana masses (these extra phases can be
measured only in $\Delta {\cal L}=2$ processes). 

Unfortunately, theoretical predictions of lepton hierarchies and
mixings depend very much on hitherto untested theoretical
assumptions. 
In the quark sector, where the bulk of the experimental data resides,
the theoretical origin of quark hierarchies and mixings is a mystery,
although there exits many theories, but none so convincing as to offer a
definitive answer to the community's satisfaction. It is therefore no 
surprise that there are  more theories of lepton
masses and mixings than there are parameters to be measured. Nevertheless, one can 
present the issues as questions:
\begin{itemize}
\item Do the right handed neutrinos have quantum numbers beyond the
standard model?
\item Are quarks and leptons related by grand unified theories?
\item Are quarks and leptons related by anomalies?
\item Are there family symmetries for quarks and leptons?
\end{itemize}

The measured numerical value of
the neutrino mass difference (barring any fortuitous degeneracies), suggests 
through the seesaw mechanism, a mass for the right-handed neutrinos
that is consistent with the scale at which
the gauge couplings unify. Is this just a numerical  coincidence, 
or should we view this as a hint for grand unification?

Grand unified theories, originally proposed as a way to treat
leptons and quarks on the same footing,  imply  symmetries much larger than 
the standard model's. Implementation of these
ideas necessitates a desert and supersymmetry, but also a carefully designed
contingent of Higgs particles to achieve the desired symmetry
breaking. That such models can be built is perhaps more of a testimony
to the cleverness of theorists rather than of Nature's. Indeed with the
advent of string theory, we know that the best features of grand
unified theories can be preserved, as most of the symmetry breaking is
achieved by geometric compactification from higher dimensions~\cite{CANDELAS}.

An alternative point of view is that the vanishing of 
chiral anomalies is necessary for consistent  theories,
and their cancellation is most easily achieved by assembling matter in
representations of anomaly-free groups. Perhaps anomaly cancellation
is more important than group structure.

Below, we present two theoretical frameworks of our work, in which one deduces
the lepton mixing parameters and masses. One is ancient~\cite{HRR},
uses the standard techniques of grand unification, but it had the 
virtue of {\it predicting} the large $\nu_\mu-\nu_\tau$ mixing
observed by SuperKamiokande. The other~\cite{ILR} is more recent, and uses
extra Abelian family symmetries to explain both quark and lepton
hierarchies. It also predicted large $\nu_\mu-\nu_\tau$ mixing, while both
schemes predict small $\nu_e-\nu_\mu$ mixings.

\subsection{A Grand Unified Model}  
The seesaw mechanism was born in the context of the grand unified 
group $SO(10)$, which naturally contains electroweak neutral 
right-handed neutrinos. Each standard model family appears  in 
two irreducible representations of $SU(5)$. However, the predictions of this
theory for Yukawa couplings is not so clear cut, and to reproduce the
known quark and charged lepton hierarchies, a special but simple set of Higgs
particles had to be included. In the simple scheme proposed by Georgi
and Jarlskog~\cite{GJ}, the ratios between the charged leptons and quark masses
is reproduced, albeit not naturally since two Yukawa couplings, not
fixed by group theory, had to be set equal. This motivated us to
generalize~\cite{HRR} their scheme to $SO(10)$, where it is 
(technically) natural, which meant that we had an automatic window
into neutrino masses through the seesaw. The Yukawa couplings were of
the Higgs-heavy,  with ${\bf 126}$ representations, but the attitude at the time 
was ``damn the Higgs torpedoes, and see what happens".  A modern
treatment would include non-renormalizable operators~\cite{BPW}, but
with similar conclusion. The model yielded the mass relations  
\be m_d-m_s=3(m_e-m_\mu)\ ;\qquad m_dm_s=m_em_\mu\ ;\ee
as well as 
\be m_b=m_\tau\ ,\ee
and mixing angles
\be
V_{us}=\tan\theta_c=\sqrt{m_d\over m_s}\ ;\qquad V_{cb}=\sqrt{m_c\over
m_t}\ .\ee
While reproducing the well-known lepton and quark mass hierarchies, it
predicted a long-lived $b$ quark, contrary to the lore of the time.
It also made predictions in the lepton sector, namely
{\bf maximal} $\nu_\tau-\nu_\mu$ mixing, small $\nu_e-\nu_\mu$
mixing of the order of $(m_e/m_\mu)^{1/2}$, and no $\nu_e-\nu_\tau$
mixing. 

The neutral lepton masses came out to be hierarchical, but heavily
dependent on the masses of the right-handed neutrinos. The
 electron neutrino mass came out much lighter
than those of $\nu_\mu$ and $\nu_\tau$. Their numerical values
depended on the top quark mass, which was then supposed to be in the
tens of GeVs!

Given the present knowledge, some of the features are remarkable, such as
the long-lived $b$ quark and the maximal $\nu_\tau-\nu_\mu$
mixing. On the other hand, the actual numerical value of the $b$ lifetime was off
a bit, and the $\nu_e-\nu_\mu$ mixing was too large to reproduce the small angle
MSW solution of the solar neutrino problem. 

The lesson should be that the simplest $SO(10)$ model 
that fits the observed quark and charged lepton hierarchies,
reproduces, at least qualitatively, the maximal mixing found by
SuperK, and predicts small mixing with the electron neutrino~\cite{CASE}.

\subsection{A Non-grand-unified  Model}
There is another way to generate hierarchies, based on adding extra
family symmetries to the standard model, without invoking grand
unification. These types of models address only the Cabibbo
suppression of the Yukawa couplings, and are not as predictive as
specific grand unified models. Still, they predict no Cabibbo
suppression between the muon and tau neutrinos. Below, we present a
pre-SuperK  model~\cite{ILR} with those features. 

The Cabibbo supression is assumed to be an indication of extra
family symmetries in the standard model. The idea is that any standard model-invariant
operator, such as ${\bf Q}_i{\bf \overline d}_jH_d$,  cannot be present
at tree-level if there are additional symmetries under which the
operator is not invariant. Simplest is to assume an Abelian symmetry,
with an electroweak singlet field $\theta$,  as its order parameter.
Then  the interaction
\be
{\bf Q}_i{\bf \overline d}_jH_d\left({\theta\over M}\right)^{n_{ij}}\ee
can appear in the potential as long as the family charges balance under the
new symmetry. As $\theta$ acquires a $vev$, this leads to a
suppression of the Yukawa couplings of the order of $\lambda^{n_{ij}}$
for each matrix element, with 
$\lambda=\theta/M$ identified with  the Cabibbo angle, and
$M$ is the natural cut-off of the effective low energy  theory. 
As a consequence of the charge balance equation
\be X_{if}^{[d]}+n^{}_{ij}X^{}_\theta=0\ ,\ee
the exponents of the suppression are related to the charge of the
standard model-invariant operator~\cite{FN},  the sum of the
charges of the fields that make up the the invariant. 

This simple Ansatz, together with the seesaw mechanism, 
implies that the family structure of the neutrino mass matrix is
determined by the charges of the left-handed lepton doublet fields. 

Each charged lepton Yukawa coupling 
$L_i\overline N_j H_u$, has an extra  charge $X_{L_i}+X_{Nj}+X_{H}$, which
gives the Cabibbo suppression of the $ij$ matrix element. Hence, 
 the orders of magnitude of these couplings can be expressed as 
\be
\pmatrix{\lambda^{l_1}&0&0\cr
0&\lambda^{l_2}&0\cr
0&0&\lambda^{l_3}\cr}{\hat Y}\pmatrix{\lambda^{p_1}&0&0\cr
0&\lambda^{p_2}&0\cr
0&0&\lambda^{p_3}\cr}\ ,\ee
where ${\hat Y}$ is a Yukawa matrix with no Cabibbo
suppressions, $l_i=X_{L_i}/X_\theta$ are the charges of the
left-handed doublets, and 
$p_i=X_{N_i}/X_\theta$, those of the singlets. The first matrix forms half of
the MNS matrix. Similarly, the mass matrix for the right-handed
neutrinos, $\overline N_i\overline N_j$ will be written in the form
\be
\pmatrix{\lambda^{p_1}&0&0\cr
0&\lambda^{p_2}&0\cr
0&0&\lambda^{p_3}\cr}{\cal M}\pmatrix{\lambda^{p_1}&0&0\cr
0&\lambda^{p_2}&0\cr
0&0&\lambda^{p_3}\cr}\ .\ee
The diagonalization of the  seesaw matrix is of the form 
\be 
L_iH_u\overline N_j \left({1\over{{\overline N}~\overline
N}}\right)_{jk}\overline N_kH_uL_l\ ,\ee
from which the Cabibbo suppression matrix from the $\overline N_i$
fields {\it cancels}, leaving us with
\be
 \pmatrix{\lambda^{l_1}&0&0\cr
0&\lambda^{l_2}&0\cr
0&0&\lambda^{l_3}\cr}\hat{\cal M}\pmatrix{\lambda^{l_1}&0&0\cr
0&\lambda^{l_2}&0\cr
0&0&\lambda^{l_3}\cr}\ ,\ee
where $\hat{\cal M}$ is a matrix with no Cabibbo suppressions.  
The Cabibbo structure of the seesaw neutrino matrix is determined
solely by the charges of the lepton doublets! As a result, the Cabibbo
structure of the MNS
mixing matrix is also due entirely to the charges of the three lepton
doublets. This general conclusion depends on the existence of at least
one Abelian family symmetry, which we argue is implied by the observed
structure in the quark sector.

The Wolfenstein parametrization of the CKM matrix~\cite{WOLF}, 
\be
\hskip 1in
\pmatrix{1&\lambda & \lambda ^3\cr
             \lambda &1&\lambda ^2\cr 
             \lambda ^3&\lambda ^2&1}\ ,\ee
and the Cabibbo structure of the quark mass ratios
\be {m_{u}\over m_t}\sim \lambda ^8\;\;\;{m_c\over m_t}\sim 
\lambda ^4\;\;\; ;
\;\;\; {m_d\over m_b}\sim \lambda ^4\;\;\;{m_s\over m_b}\sim
\lambda^2\ ,\ee
can be reproduced~\cite{ILR,EIR} by a simple {\it family-traceless} charge assignment 
for the three quark families, namely
\be
X_{{\bf Q},{\bf \overline u},{\bf \overline d}} ={\cal B}(2,-1,-1)+
\eta_{{\bf Q},{\bf \overline u},{\bf \overline d}}(1,0,-1)\ ,\ee
where ${\cal B}$ is baryon number, $\eta_{{\bf \overline d}}=0 $, and 
$\eta_{{\bf Q}}=\eta_{{\bf \overline u}}=2$. 
Two striking facts are evident: 
\begin{itemize}
\item the charges of the down quarks, ${\bf \overline d}$, 
associated with the second and third families are the same, 
\item ${\bf Q}$ and ${\bf \overline u}$ have the same value for 
$\eta$.
\end{itemize}
To relate these quark charge assignments to those of the leptons,
we need to inject some more theoretical prejudices. Assume these
 family-traceless charges are gauged, and not anomalous. Then to
cancel anomalies, the leptons must themselves have family charges. 

Anomaly cancellation generically implies group structure. In $SO(10)$, 
baryon number generalizes to ${\cal B}-{\cal L}$, where ${\cal L}$ is total
lepton number, and  in 
$SU(5)$   the fermion assignment is ${\bf\overline 5}={\bf
\overline d}+L$, and ${\bf 10}={\bf Q}+{\bf \overline u}+\overline
e$. Thus anomaly cancellation is easily achieved by
assigning $\eta=0$ to the lepton doublet $L_i$, and $\eta=2$ to the
electron singlet $\overline e_i$, and by generalizing baryon number to
${\cal B}-{\cal L}$, leading to the charges
\be
X_{{\bf Q},{\bf \overline u},{\bf \overline d}, L,\overline e} =({\cal
B}-{\cal L})(2,-1,-1)+
\eta_{{\bf Q},{\bf \overline u},{\bf \overline d}}(1,0,-1)\ ,\ee
where now $\eta_{{\bf \overline d}}=\eta_{L}=0 $, and 
$\eta_{{\bf Q}}=\eta_{{\bf \overline u}}=\eta_{\overline e}=2$. 

The charges of the
lepton doublets are simply $X_{L_i}=-(2,-1,-1)$. We have just
argued that these charges determine the Cabibbo structure of the MNS
lepton mixing matrix to be
\be
\hskip .2in
{\cal U}^{}_{MNS}\sim\pmatrix{1&\lambda^3&\lambda^3\cr
\lambda^3&1&1\cr \lambda^3&1&1\cr}\ ,\ee
implying {\it  no Cabibbo suppression in the mixing between
$\nu_\mu$ and $\nu_\tau$}. This is consistent with the SuperK
discovery and  with the small angle MSW~\cite{MSW} solution to the solar
neutrino deficit. One also obtains a much lighter electron neutrino, and
Cabibbo-comparable masses for the muon and tau neutrinos. Notice that
these predictions are  subtly different from those
of grand unification, as they  yield $\nu_e-\nu_\tau$ mixing. 
It also implies a much lighter electron neutrino, and
Cabibbo-comparable masses for the muon and tau neutrinos. 

On the other hand, the scale of the neutrino mass values depend on the family 
trace of the family charge(s). Here we simply quote the results
our model~\cite{ILR}. The masses of the right-handed neutrinos are found to be of the
following orders of magnitude
\be
m_{\overline N_e}\sim M\lambda^{13}\ ;\qquad m_{\overline N_\mu}\sim
m_{\overline N_\tau}\sim M\lambda^7\ ,\ee
where $M$ is the scale of the right-handed neutrino mass terms,
assumed to be the cut-off. The seesaw mass matrix for the three light  neutrinos 
comes out to be 
\be
\hskip .5in
 m^{}_0\pmatrix{a\lambda^6&b\lambda^3&c\lambda^3\cr
b\lambda^3&d&e\cr
c\lambda^3&e&f\cr}\ ,\ee
where we have added for future reference the prefactors $a,b,c,d,e,f$, all of 
order one, and 
\be m_0^{}={v_u^2\over
{M\lambda^3}}\ ,\ee
where $v_u$ is the $vev$ of the Higgs doublet. This matrix has one light eigenvalue
\be
m_{\nu_e}\sim m_0^{}\lambda^6_{}\ .\ee
Without a detailed analysis of the prefactors, the masses of the other 
two neutrinos come out  to be both of 
 order $m_0$. 
The mass difference announced by  superK~\cite{SUPERK}  cannot  be
reproduced without going beyond the model, by taking into account the
prefactors. The two heavier mass
eigenstates and their mixing angle are written in terms of 
\be
x={df-e^2\over (d+f)^2}\ ,\qquad y={d-f\over d+f}\ ,\ee
as
\be {m_{\nu_2}\over m_{\nu_3}}={1-\sqrt{1-4x}\over 1+\sqrt{1-4x}}\
,\qquad \sin^22\theta_{\mu\tau}=1-{y^2\over 1-4x}\ .\ee
If $4x\sim 1$, the two heaviest neutrinos are nearly degenerate. If
$4x\ll 1$, a condition easy to achieve if $d$ and $f$ have the same
sign, we can obtain an adequate split between the two mass
eigenstates. For illustrative purposes, when $0.03<x<0.15$, we find
\be
4.4\times 10^{-6}\le \Delta m^2_{\nu_e-\nu_\mu}\le 10^{-5}~{\rm eV}^2\
  ,\ee
which yields the correct non-adiabatic MSW~\cite{MSW} effect, and
\be
5\times 10^{-4}\le  \Delta m^2_{\nu_\mu-\nu_\tau}\le 5\times
10^{-3}~{\rm eV}^2\ ,\ee
for the atmospheric neutrino effect. These were calculated with a
cut-off, $10^{16}~{\rm GeV}<M<4\times 10^{17}~{\rm GeV}$, and a mixing
angle, $0.9<\sin^22\theta_{\mu-\tau}<1$. This value of the cut-off is 
 compatible not only with the data but also with the gauge
coupling unification scale, a necessary condition for the consistency
of our model, and more generally for  the basic ideas of grand unification. 

\subsection{Proton Decay}
We have seen in the previous section that the ultraviolet cut-off $M$ 
appears directly in the seesaw masses. Now that it is determined by 
experiment, we can use it to estimate the strength of other
interactions, in particular those that  generate proton decay. 
In  a supersymmetric theory with no R-parity 
violation, proton decay is caused by two types of operators that
appear in the superpotential as

\be W={1\over M}[{\kappa _{112i}}{\bf Q}_1{\bf Q}_1{\bf Q}_2{\bf
L}_i+{\overline \kappa}_{1jkl}
{\bf {\overline u}}_1{\bf {\overline u}}_j{\bf {\overline d}}_k{\bf {\overline 
e}}_l]\ee
where for the first operator the flavor index $i=1,2$ if there is a charged 
lepton in
the final state and $i=1,2,3$ if there is a neutrino and $j=2,3$,
$k,l=1,2$. Operators that involve only one family, such as ${\bf Q}_1{\bf Q}_1{\bf Q}_1{\bf L}_i$, and ${\bf {\overline u}}_1{\bf {\overline u}}_1{\bf {\overline d}}_1{\bf {\overline e}}_l$ are forbidden by symmetry.  The reasons are that  the combination
 ${\bf Q}_1{\bf Q}_1{\bf Q}_1$ vanishes identically in the color singlet channel, and the combination ${\bf {\overline u}}_1{\bf {\overline u}}_1$ transforms as a color sextet, and cannot make a color invariant with the addition of an extra antiquark. 
This is the well-known statement that in supersymmetric theories,
proton decay products will necessarily involve strange particles. The
conventional decay into first family members is still there but not
dominant. It would be most amusing if the first experimental manifestation of
supersymmetry were to be the detection of proton decay into kaons!

These interactions lead to dimension-five four-body interactions between two squarks and two sparticles (two squarks or two sleptons). After  gaugino exchange, the two sparticles are turned into particles~\cite{protondecay}, leading to baryon number viola
ting four fermion interactions, among them  proton decay. The existing bounds on proton decay put severe constraints on the couplings ${\kappa _{112i}}$, and ${\overline \kappa}_{1jkl}$.

In theories where the Cabibbo suppression of operators is related to their charges, we expect these operators to be highly Cabibbo-suppressed. This is because of sum rules which relate their charges to those of standard model invariants. 

Under the assumptions of  tree-level top quark mass, zero $\mu$-term charge, and of the Green-Schwarz relation $C_{\rm color}=C_{\rm weak}$, the family-independent charges  satisfy 
\be X_{{\bf Q}_1{\bf Q}_1{\bf Q}_2{\bf
L}_i}=X_{{\bf {\overline u}}_1{\bf {\overline u}}_j{\bf {\overline d}}_k{\bf {\overline 
e}}_l}=X_{{\bf Q}_1{\bf {\overline u}}_1H_u}\ .\ee 

Also, the branching ratios between different proton decay modes are determined by the $U(1)$ charges that are flavor dependent. In our model~\cite{ILR}, the least suppressed operator is $ {\bf Q}_1{\bf Q}_1{\bf Q}_2{\bf L}_{2,3}$, with
\be
{\kappa _{1122}}\sim {\kappa _{1123}}\sim\lambda^{11}\ ,\ee
leading to the estimate (with $M$ set by the neutrino mass values), 
\be \Gamma(p\to K^0+\mu^+)\sim 10^{32} ~{\rm yr^{-1}}\ , \ee
at the same level as the SuperK limits presented at this workshop by L. Sulak.

It is unfortunate that these models yield only orders of magnitude estimate, but it should be clear that those decay rates are tantalizingly close to the experimental bounds. Thus it is important to build a larger proton decay detector and improve the bou
nds by at least one order of magnitude.

\section{Outlook}
Theoretical predictions of neutrino masses and mixings depend on developing
a credible theory of flavor. We have presented
two flavor schemes, which predicted not only maximal $\nu_\mu-\nu_\tau$
mixing, but also small $\nu_e-\nu_\mu$ mixings. Neither scheme includes
sterile neutrinos~\cite{dienes}. The present experimental situation is somewhat
unclear: the LSND results~\cite{LSND} imply the presence of a sterile
neutrino; and  superK favors 
$\nu_\mu-\nu_\tau$ oscillation over $\nu_\mu-\nu_{\rm sterile}$. The origin
of the solar neutrino deficit remains  a puzzle, which several
possible explanations. One is the non-adiabatic MSW effect in
the Sun, which our theoretical ideas seem to favor, but it is an
experimental question which is soon to be answered by the
continuing monitoring of the $^8B$ spectrum by SuperK, and the advent  
of the SNO detector. If neutrino masses  reflect (through the seesaw) the value of the ultraviolet cut-off, they set the scale for the strength of proton decay interactions, implying that observation may not be far in the future. Neutrino physics has give
n us a first glimpse of physics at very short distances, and proton decay cannot be too far behind. 

\section{Acknowledgements}
I wish to thank Professors  C. K. Jung and M. V. Diwan  for inviting
me to this important and very stimulating workshop. This research was 
supported in part by the department
of energy under grant DE-FG02-97ER41029.

\end{document}